\documentclass[amsmath,amssymb,aps,pre,twocolumn,longbibliography]{revtex4-1}
\usepackage{physics}

\usepackage{graphicx}
\usepackage{bm}
\usepackage{hyperref}

\AtBeginDocument{%
    \newwrite\bibnotes
    \def\bibnotesext{Notes.bib}
    \immediate\openout\bibnotes=\jobname\bibnotesext
    \immediate\write\bibnotes{@CONTROL{REVTEX41Control}}
    \immediate\write\bibnotes{@CONTROL{%
    apsrev41Control,author="08",editor="1",pages="1",title="0",year="1"}}
     \if@filesw
     \immediate\write\@auxout{\string\citation{apsrev41Control}}%
    \fi
}%

\begin{document}
	
\title{On the possibility of exploring tip-molecule interactions\\ with STM experiments}

\author{Christoph Schiel}
\author{Philipp Rahe}
\author{Philipp Maass}
\affiliation{Fachbereich Physik, Universit\"at Osnabr\"uck, Germany}
\email{maass@uni-osnabrueck.de}

\date{August 1, 2022}

\begin{abstract}
We present a theory for analyzing residence times of single molecules
in a fixed detection area of a scanning tunneling microscope
(STM). The approach is developed for one-dimensional molecule
diffusion and can be extended to two dimensions by using the same
methodology. Explicit results are derived for an harmonic attractive
and repulsive tip-molecule interaction. Applications of the theory
allows one to estimate the type and strength of interactions between
the STM tip and the molecule. This includes the possibility of an
estimation of molecule-molecule interaction when the tip is decorated
by a molecule. Despite our focus on STM, this theory can analogously
be applied to other experimental probes that monitor single molecules.
\end{abstract}

\maketitle´

\section{Introduction}
For controlling the kinetic growth and self-assembly of molecules on
surfaces, the molecular diffusion coefficient is one important
parameter \cite{Rahe/etal:2013}. Multiple methods have been developed to
obtain this parameter \cite{Barth:2000}. Here we focus on a specific
technique of determining the diffusion coefficient from the analysis
of current fluctuations in scanning tunneling microscopy (STM). These
fluctuations are caused by molecules entering and leaving the
detection area under an STM tip at a fixed spatial position
\cite{Lozano/Tringides:1995, Ikonomov/etal:2010, Hahne/etal:2013, Hahne/etal:2014}. Thus, this
method is particularly well suited for molecules with high mobility
since the tunneling current can be measured at a high time
resolution. It can also be applied in other fields as, for example,
ﬂuorescence correlation spectroscopy \cite{Zumofen/etal:2004, Petrov/Schwille:2008,
  Koynov/Butt:2012, Motegi/etal:2013, Steger/etal:2013}.
 
\begin{figure}[b!]
\centering
\includegraphics[width=0.7\columnwidth]{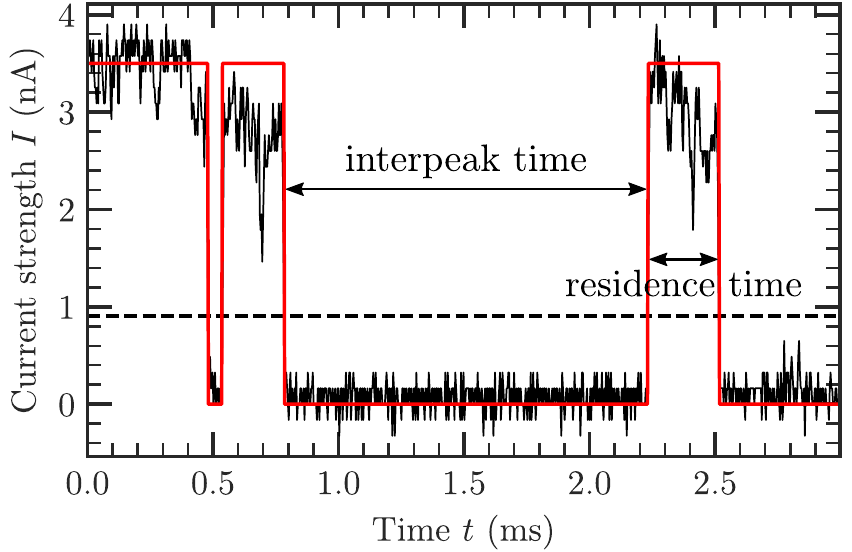}
\caption{Typical tunneling current measured with an STM. In this
  example, the fluctuations by diffusing PTCDA
  (perylenetetracarboxylic dianhydride) molecules on the Ag(110)
  surface are sampled.  The black line is the measured current. The
  dotted line indicates the threshold value. Currents above this
  threshold are caused by the presence of PTCDA molecules under the
  tip. The red line indicates the resulting dichotomic fluctuations of
  the current (binary signal). The width of a rectangular peak in the
  binary representation gives the residence time. The time between the
  termination and start of two consecutive rectangular peaks gives the
  interpeak time.}
\label{fig:illustration_current_analysis}
\end{figure}

In previous studies, different techniques have been developed and
investigated to evaluate the times series of the tunneling current
\cite{Hahne/etal:2013,Hahne/etal:2014}. These studies include molecules of
different shapes, as well as effects of additional rotational
diffusion. With an appropriate analysis, the time series of the
tunneling current can be mapped onto a binary signal (dichotomic
fluctuations), where the signal value is one for a molecule under the
tip and zero otherwise, see Fig.~\ref{fig:illustration_current_analysis} for an
example of the resulting rectangular peaks.

The stochastic behavior of the binary signal can be characterized by
two characteristic times, namely the residence time and the interpeak
time. The residence time is the duration of a rectangular peak and the
interpeak time is the time interval between two consecutive
rectangular peaks. By sampling these times in a long time series, a
residence time distribution (RTD) and an interpeak time distribution
(ITD) are obtained. Based on theoretical modeling without tip-molecule
interactions, these measured distributions can be fitted to analytical
predictions and diffusion coefficients are extracted as fitting
parameters. As a third approach, the experimentally determined
autocorrelation function (ACF) of the signal is compared with
theoretical expressions.

The ITD is governed by the diffusive dynamics of the molecules while
they are outside the detection area of the tip. Accordingly,
tip-molecule interactions does not influence the diffusion coefficient
under the assumption that the tip-molecule interaction range is
shorter or equal to the detection range. This is in marked contrast to
the RTD, where the diffusive dynamics is examined while a molecule is
located in the tip-sample gap. This contrasting behavior can be
exploited to probe tip-molecule interactions by modeling the influence
of these interactions on the diffusion coefficient. This could become
an interesting new technique for analyzing molecule-molecule
interactions beyond current methods such as atomic force microscopy,
in particular in view of the possibility to decorate STM tips with
specific molecules. Comparison of ITDs and RDTs measured in one
experiment can thus become a valuable tool for analyzing
molecule-molecule interactions. The ACF approach would be not suitable
for such analysis because both the diffusion inside and outside the
detection area under the tip enter its functional behavior.

Here, we perform first steps to evaluate this idea of quantifying
tip-molecule interactions via their influence on diffusion
coefficients. To this end, we present a theory for the RTD in the
presence of a harmonic interaction between tip and molecule. Using
this theory, we evaluate how such interactions modify the RTD for a
freely diffusing molecule on the surface. This includes both
attractive and repulsive interactions with the tip.

\section{Residence time distribution in the presence of tip-molecule interaction: General formal solution}
\label{subsec:tipinf}

We focus on a one-dimensional diffusion of the molecules on the
surface, which allows us to obtain explicit analytical expressions. A
one-dimensional treatment can also be a appropriate to describe
experimental situations of highly anisotropic molecule diffusion, as
it is encountered, for example, on (110) surfaces.

To be specific, we consider the diffusion of a molecule at a
temperature $T$ under a tip pointing to the origin of the coordinate
frame. The tip-molecule interaction is described by $V(x) = V(-x)$,
where $x$ is the center of the molecule. The detection interval of the
molecule under the tip is $[-L/2,L/2]$, corresponding to a spacial
tip-molecule interaction range of $L/2$. In the absence of
tip-molecule interaction ($V(x)=0$), the molecule diffusion
coefficient is $D_0$.

The residence time is the time between the event where a molecule
center enters the detection interval and the event where it leaves
that interval thereafter. Accordingly, the RTD is equal to a
first-passage time distribution. This distribution follows from the
probability $p(x,t) {\rm d}x$ of finding the molecule in the interval
$[x,x+{\rm d}x]$ under the conditions that absorbing boundaries are
present at $-L/2$ and $L/2$,
\begin{equation}
p(-L/2,t) = p(L/2,t) = 0\,.
\label{eq:boundary-condition}
\end{equation}
The initial molecule position $p(x,0)$ at time $t=0$ is $-L/2 +
\epsilon$ (or $L/2 - \epsilon$) where $\epsilon$ is a small
displacement away from the absorbing boundary. Accordingly, we can set
\begin{equation}
\label{eq:pzero}
p(x,0) = \frac{1}{2} \left[\delta\big(x+(L/2-\epsilon)\big) + \delta(x-\big(L/2-\epsilon)\big)\right]\,.
\end{equation}
We take $\epsilon$ as the root mean squared displacement of a freely
diffusing molecule within the time resolution interval $\Delta t$ of
the time series, $\epsilon= \sqrt{D_0\Delta t}$. The first-passage
time distribution or RTD is given by \cite{Redner:2001}
\begin{equation}
\label{eq:rdt}
\Psi(t) = -\frac{\partial}{\partial t} \int_{-L/2}^{L/2}p(x,t)\,,
\end{equation}
where $p(x,t)$ is the probability density for the initial condition
\eqref{eq:pzero} and absorbing boundary condition
\eqref{eq:boundary-condition}.  Its time evolution is given by the
diffusion equation \cite{Risken:1985}
\begin{equation}
\label{eq:diffusionequation}
\frac{\partial p(x,t)}{\partial t} = L_s(x) p(x,t)\,,
\end{equation}
where
\begin{equation}
L_s(x) = D_0 \left[\frac{\partial^2}{\partial x^2}-\beta\frac{\partial}{\partial x} V'(x)\right]
\end{equation}
is the Smoluchowski time evolution operator; 
$\beta =1/k_{\scriptscriptstyle\text{B}}T$ is the inverse thermal energy.

A general formal solution of Eq.~\eqref{eq:diffusionequation} can
be obtained by introducing the self-adjoint operator
\cite{Risken:1985},
\begin{align}
L(x)&= e^{\beta V(x) / 2}L_s(x) e^{-\beta V(x) / 2}\nonumber\\
&= D_0 \left[\frac{\partial^2}{\partial x^2} + \frac{\beta}{2} V''(x) - \frac{\beta^2}{4}V'(x)^2\right]\,,
\end{align}
which is the time evolution operator for the rescaled probability density $e^{\beta V(x)/2}p(x,t)$. 
The solution for $p(x,t)$ is 
\begin{equation}
\label{eq:solutionp}
p(x,t) = e^{-\beta V(x)/2} \sum_{n}a_n\varphi_n(x)e^{-\lambda_nt}
\end{equation}
where
\begin{equation}
\label{eq:an}
a_n = \int_{-L/2}^{L/2}\varphi_n(x)e^{\beta V(x)/2} p(x,0) \mathrm{d}x
\end{equation}
and $\varphi_n(x)$ are the eigenfunctions of the operator $L(x)$,
\begin{equation}
\label{eq:eigenvalueproblem}
L(x)\varphi_n(x) = -\lambda_n \varphi_n(x)\,.
\end{equation}
The eigenfunctions must satisfy the same boundary conditions
\eqref{eq:boundary-condition} as $p(x,t)$,
\begin{equation}
\label{eq:bound}
\varphi(-L/2) = \varphi(L/2) = 0\,.
\end{equation}
Inserting $p(x,t)$ from Eq.~\eqref{eq:solutionp} in Eq.~\eqref{eq:rdt}
gives
\begin{equation}
\label{eq:RTD}
\Psi(t) = \sum_{n} a_n b_n \lambda_n e^{-\lambda_n t}
\end{equation}
where
\begin{equation}
b_n = \int_{L/2}^{L/2}\mathrm{d}x\,\varphi_n e^{-\beta V/2}\,.
\end{equation}
Using this approach, the determination of the RTD reduces to a
calculation of the eigenfunctions $\varphi(x)$.

In the non-interacting case $V(x) = 0$, Eq.~\eqref{eq:RTD} yields
\begin{equation}
\label{eq:RTD0}
\Psi_0 = \frac{4}{\pi\tau_L}\sum_{n=0}^{\infty} (2n+1)\sin\left(\frac{(2n+1)\pi\epsilon}{L}\right)e^{-(2n+1)^2t/\tau_L}\,,
\end{equation}
where
\begin{equation}
\label{eq:taul}
\tau_L = \frac{L^2}{\pi^2D_0}\,.
\end{equation}
It can be shown \cite{Hahne/etal:2013} that Eq.~\eqref{eq:RTD0} gives a
power law $\propto t^{-3/2}$ for $\Delta t \ll t \ll \tau_L$ and an
exponential decay for $t \gg \tau_L$.

\section{Residence time distribution for harmonic tip-molecule interaction}
Generally, a tip-molecule interaction can be of complex physical
origin and include numerous contributions. For simplicity, we here
consider a potential along the lateral position $x$ of the harmonic
form
\begin{equation}
V(x) = \left\{\begin{array}{cc} 
\displaystyle V_0 -\frac{k}{2}x^2\,, &x\in [-L/2,L/2],\\[2ex] 
0\,, & \text{otherwise}
\end{array}\right.
\end{equation}
with $V_0 = kL^2/8$, wich makes the potential continuous at $x=\pm
L/2$. The stiffness $k$ is positive (negative) for repulsive
(attractive) interaction.

The eigenvalue problem in Eq.~\eqref{eq:eigenvalueproblem} then reads
\begin{equation}
D_0\left[\frac{\mathrm{d}^2}{\mathrm{d}x^2}-\frac{\beta k}{2}
-\frac{\beta^2 k^2}{4}x^2\right]\varphi_n(x) = \lambda_n \varphi_n(x)
\end{equation}
in explicit form.  Introducing the dimensionless length $y= \sqrt{\pm
  \beta k}\,x$ and defining $\tilde{\varphi}_n(y) = \varphi_n(x) =
\varphi_n(y/\sqrt{\pm\beta k})$, we obtain the differential equation
for parabolic cylinder functions \cite{Abramowitz/Stegun:1965},
\begin{equation}
\label{eq:paraboliceq}
\frac{\mathrm{d}^2\tilde{\varphi}_n(y)}{\mathrm{d} y^2} - \left(\frac{1}{4}y^2 \pm c_n\right)\tilde{\varphi}_n(y) = 0\,.
\end{equation}
Here and in the following the upper (lower) sign refers to repulsive
(attractive) interaction, and
\begin{equation}
\label{eq:cn}
c_n = \frac{1}{2}-\frac{\lambda_n}{\beta k D_0}\,.
\end{equation}

Due to the symmetry of the potential, the eigenfunctions
$\tilde{\varphi}_n$ can be chosen to be eigenfunctions also of the parity
operator.  The fundamental system of solutions of
Eq.~\eqref{eq:paraboliceq} is thus given by even and odd
solutions. The even solutions are
\begin{equation}
\tilde{\varphi}_n(y) = e^{-\frac{1}{4} y^2}M\left(\frac{1}{4} \pm \frac{c_n}{2},\frac{1}{2},\frac{1}{2}y^2\right)\,,
\end{equation}
where $M(\alpha,\gamma,z)$ is Kummer's confluent hypergeometric
function.  The odd solutions do not contribute to the RTD
\eqref{eq:RTD}, because the respective $a_n$ in Eq.~\eqref{eq:an} are
zero for these odd solutions due to the symmetry of the potential and
of the initial condition \eqref{eq:pzero}.

The $c_n$ or corresponding eigenvalues $\lambda_n$
[cf.\ Eq.~\eqref{eq:cn}] follow from the boundary conditions in
Eq.~\eqref{eq:bound},
\begin{equation}
\label{eq:lambdan}
M\left(\frac{1}{2}\pm\frac{c_n}{2},\frac{1}{2},\pm\frac{\beta k L^2}{8}\right) = 0\,.
\end{equation}
The eigenfunctions are
\begin{equation}
\label{eq:eigenfunctions}
\varphi_n(x) = \mathcal{N}_n e^{\mp\beta k x^2/4} 
M\left(\frac{1}{2}\pm\frac{c_n}{2} ,\frac{1}{2},\pm\frac{\beta kx^2}{2}\right)
\end{equation}
with normalization factor
\begin{equation}
\mathcal{N}_n = \left[\int_{-L/2}^{L/2} \varphi_n(x)^2 \mathrm{d}x\right]^{-1}\,.
\end{equation}

\begin{figure}[t!]
\centering
\includegraphics[width=\columnwidth]{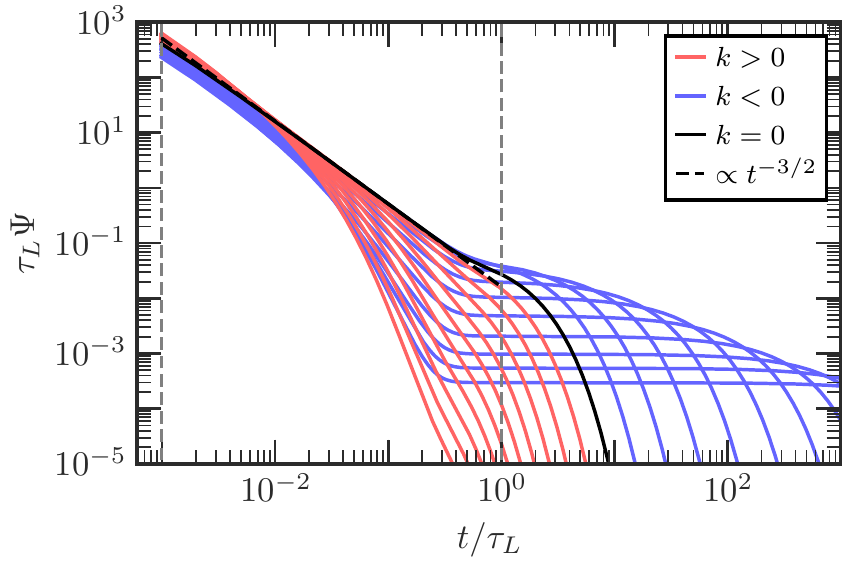}
\caption{Scaled residence time distributions $\tau_L\Psi$ for
  attractive (blue lines) and repulsive (red lines) interactions. The
  black line shows $\tau_L\Psi_0$ in the absence of tip-molecule
  interactions [see Eq.~\eqref{eq:RTD0}]. In between the dashed gray
  lines that corresponds to the time interval between $\Delta t$ (time
  resolution) and $\tau_L$ (see Eq.~\eqref{eq:taul}), $\Psi_0$ decays
  as a power law, $\Psi_0\sim t^{-3/2}$. The scaled stiffness 
  $\beta k L^2$ is varied in steps of $10$ from $\beta k L^2=-100$ to 100
  (with increasing $\beta |k| L^2$ the curves shift further away from
  the $\Psi_0(t)$ curve). The displacement $\epsilon$ entering the
  initial condition \eqref{eq:pzero} is $\epsilon = 0.01L$.}
\label{fig:RTDs_for_attractive_and_repulsive_interactions}
\end{figure}

\begin{figure*}[t!]
\centering
\includegraphics[width=0.49\textwidth]{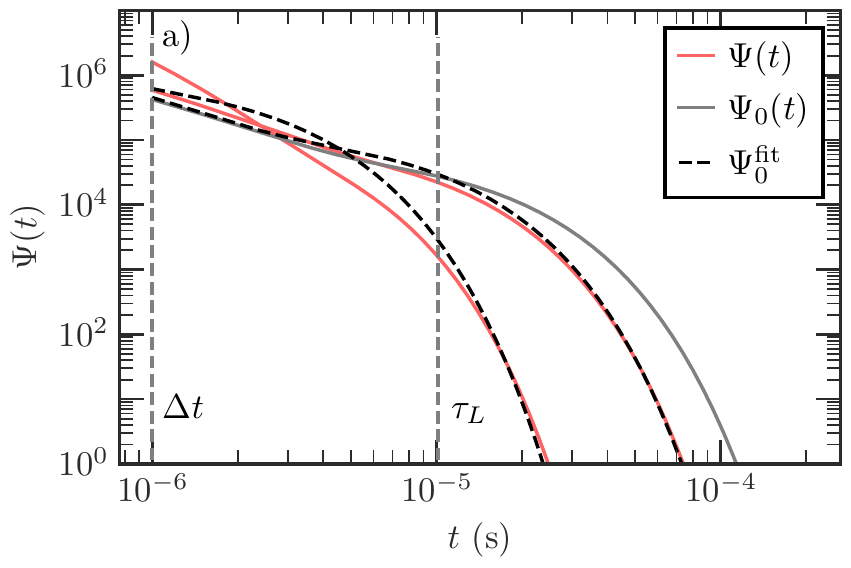}
\includegraphics[width=0.49\textwidth]{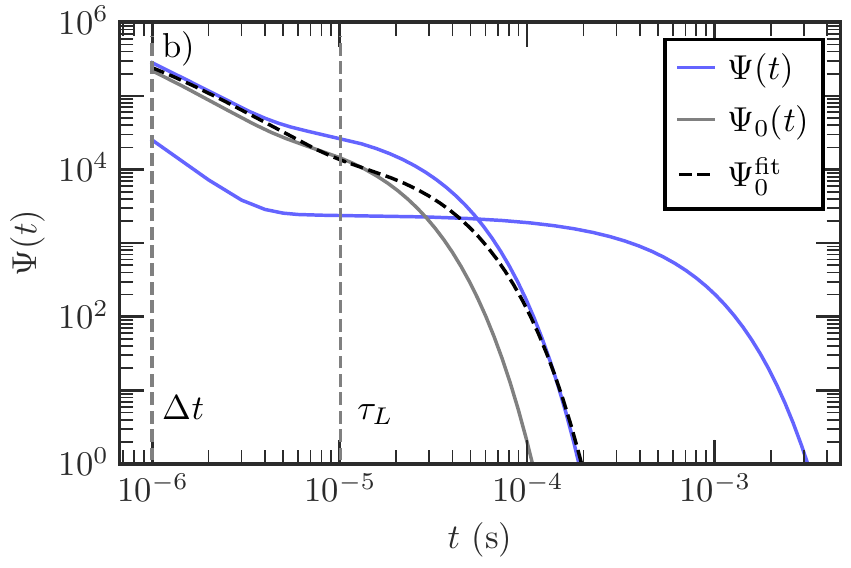}
\caption{Residence time distributions $\Psi(t)$  for (a) repulsive (red solid lines) and
  (b) attractive interactions (red solid lines) of two scaled stiffnesses 
 $\beta k L^2=\pm10$ and $\beta k L^2=\pm50 $. The gray solid lines shows the RTD
  $\Psi_0(t)$ in the absence of tip-molecule
  interaction. The curves representing $\Psi_0^\text{fit}$ (dashed black lines)
  are fits of Eq.~\eqref{eq:RTD0}
  to $\ln \Psi(t)$ with the method of least-squares, i.e.\ fits under the assumption 
  that a tip-molecule interaction is not present. The results are
  shown for typical parameters estimated from the RTD analysis
  described in Ref.~\cite{Hahne/etal:2013}: $D_0/L^2=10^{4}\,{\rm s}^{-1}$ and
  $\epsilon/L = 0.1$. The dashed gray lines mark the time interval
  between $\Delta t$ (time resolution) and $\tau_L$ [see Eq.~\eqref{eq:taul}]. This interval is narrower
   here than in Fig.~\ref{fig:RTDs_for_attractive_and_repulsive_interactions} due to the
  larger value of $\epsilon/L$.}
\label{fig:RTD_comparison_with_psi0_fits}
\end{figure*}

Figure~\ref{fig:RTDs_for_attractive_and_repulsive_interactions} shows scaled RTDs $\tau_L\Psi$ as a
function of $t/\tau_L$ for repulsive and attractive harmonic
interactions of varying stiffness $k$.
The black curve represents the RTD $\Psi_0(t)$ in the absence of
tip-molecule interaction. As mentioned above, it decay as a power law
$\sim t^{-3/2}$ for $\Delta t < t < \tau_L$ and exponentially for
$t>\tau_L$.

For analyzing the impact of the tip-molecule interaction, RTDs in
Fig~\ref{fig:RTDs_for_attractive_and_repulsive_interactions} are displayed for scaled stiffnesses 
$\beta k L^2$ varied in steps of $10$ from $\beta k L^2=-100$ to 100
($\beta k L^2 = \pm 10$ are the closest lines to the black curve
representing the RTD $\Psi_0$ in the non-interacting case). 
For small stiffness, $\beta |k| L^2\leq 10$, deviations from
$\Psi_0(t)$ are mainly present in the exponential tail of the
distribution, while they are hardly visible in the power law
regime. With increasing $\beta |k| L^2$, deviations become more pronounced
and can be seen also for small times $t < \tau_L$.

For the attractive interaction, the probability of a molecule to leave
the detection interval at small times $t \ll \tau_L$ is reduced
compared to $\Psi_0(t)$. For large times $t>\tau_L$, by contrast, it
is strongly increased. This is in agreement with the expected behavior
for attractive interaction, where the tip is tending to pull the
molecules towards the center of the detection interval.

The behavior for repulsive tip-molecule interactions is reversed: At
small times $t <\tau_L$, the interaction tends to push the molecule
out of the detection interval. The probability for the molecule to
leave this interval at small times is thus increased compared to
$\Psi_0(t)$. For larger times, $\Psi(t)$ decreases more rapidly
than $\Psi_0(t)$, leading to significantly shorter mean residence
times.

For both attractive and repulsive tip-molecule interaction, the power
law ceases to be present at larger stiffness $|k|$ and the overall
behavior can be effectively characterized by a single exponential
decay.

\section{Implications for estimating molecule diffusion coefficients and tip-molecule interaction}
In this section, we first assess when a
difference of RTD data is apparent for the cases with and without interaction.
Secondly, we clarify whether a corresponding difference
can be observed experimentally. 

We start by evaluating
the impact on diffusion parameters when neglecting the presence of
the tip-molecule interactions.
As for the parameters $D_0$ and $\epsilon$ in the model, 
we use typical values obtained in the RTD analysis of
Ref.~\onlinecite{Hahne/etal:2013}. There, the RTD analysis of freely moving
copper phthalocyanine molecules on the Ag(100) surface yielded $D_0
\simeq 10^{-10} \text{cm}^2/\text{s}$ at a temperature of 222~K and
$L$ in the nm regime. The time resolution $\Delta t$ was of the order
of $\mu\rm s$, resulting in $\epsilon \approx 0.1L$. 

Taking $L=1 \text{nm}$,  and setting $D_0/L^2=10^4\,{\rm s}^{-1}$ and  $\epsilon = 0.1L$,
the eigenvalues $\lambda_n$ and eigenfunctions $\varphi_n(x)$ are
determined from Eq.~\eqref{eq:lambdan} and
Eq.~\eqref{eq:eigenfunctions}, respectively.  Using Eq.~\eqref{eq:RTD},
RTDs for repulsive
and attractive tip-molecule interactions are calculated for the two
scaled stiffness values $\beta |k| L^2 = 10$ and $50$, with the
coefficients $a_n$ and $b_n$ in Eq.~\eqref{eq:RTD} obtained from
Eqs.~\eqref{eq:lambdan} and \eqref{eq:eigenfunctions}.

The corresponding RTDs $\Psi(t)$ are shown in
Fig.~\ref{fig:RTD_comparison_with_psi0_fits} for (a) $\beta k  L^2= 10$
and $50$ (red lines), and (b) 
$\beta k  L^2=-10$ and $-50$ (blue lines). For
comparison, the RTD $\Psi_0(t)$ for the non-interacting case ($k=0$)
is shown in both figures (gray lines). 
Clear differences between
$\Psi_0(t)$ and $\Psi(t)$ are apparent for these realistic
parameters. 

Let us consider the red curves as representing
experimental results that would have been measured in the presence of
tip-molecule interactions. For evaluating the influence of
these interactions on the RTD, however, the analysis is first carried out under the assumption that the
tip-molecule interactions are not present. Accordingly, the expression given in Eq.~\eqref{eq:RTD0} is
fitted to the red curves, where only the two parameters $\tau_L =
L^2/\pi^2D_0$ and $\tilde{\epsilon}=\epsilon/L$ are adjusted.

We determined optimized values $\tau_L^{\scriptscriptstyle{\rm
    fit}}$ and $\tilde{\epsilon}^{\scriptscriptstyle{\rm fit}}$ by
applying the method of least-squares to $\ln\Psi(t)$. The logarithm is taken because the
characteristic changes of $\Psi(t)$ occur on scales covering several
orders of magnitude, see Figs.~\ref{fig:RTDs_for_attractive_and_repulsive_interactions} and \ref{fig:RTD_comparison_with_psi0_fits}. For both
fit parameters, we specified ranges $\tau_L^\text{min} < \tau_L^{\scriptscriptstyle{\rm fit}} <
\tau_L^\text{max}$ and $\tilde{\epsilon}_\text{min} < \tilde{\epsilon}^{\scriptscriptstyle{\rm fit}}
< \tilde{\epsilon}_\text{max}$. In a self-consistency check, we
verified that the fit of Eq.~\eqref{eq:RTD0} to $\ln \Psi_0(t)$ yields
exactly the input parameters, i.e.\ $\tau_L^{\scriptscriptstyle{\rm
    fit}} = \tau_L$ and $ \tilde{\epsilon}^{\scriptscriptstyle{\rm
    fit}} = \epsilon/L$. 
    The black dashed lines in
Fig.~\ref{fig:RTD_comparison_with_psi0_fits} are corresponding fits $\Psi_0^{\rm fit}$ of Eq.~\eqref{eq:RTD0},
i.e., under neglect of tip-molecule interaction,
to the red lines representing experimental RTDs $\Psi(t)$
under influence of tip-molecule interaction.

For the repulsive interaction, we find that the fitting yields
$\tilde{\epsilon}^{\scriptscriptstyle{\rm fit}} =
\tilde{\epsilon}_\text{min}$, even when further reducing the lower limit
$\tilde{\epsilon}_\text{min}$ of the fitting range for $\tilde{\epsilon}^{\scriptscriptstyle{\rm fit}}$. 
This is due to the fact that the fit
under neglect of tip-molecule interaction 
underestimates $\Psi(t)$ for small times, as discussed above in
connection with Fig.~\ref{fig:RTDs_for_attractive_and_repulsive_interactions}(a). This
underestimation becomes less pronounced for smaller
$\epsilon$. However, the reduction of the underestimation upon
decreasing $\tilde{\epsilon}_\text{min}$ is very weak and would
hardly be noticeable when analysing experimental data
affected by noise. 

For the further analysis, we take
$\tilde{\epsilon}_\text{min}=\tilde{\epsilon}^{\scriptscriptstyle{\rm
    fit}} = 0.1$. We then obtain $\tau_L^{\scriptscriptstyle{\rm fit}} = 6.1\times
10^{-6}\,\rm s$ and $\tau_L^{\scriptscriptstyle{\rm fit}} = 1.7\times
10^{-6}\, \rm s$ for $\beta k L^2 = 10$ and 
50, respectively. The dotted lines in
Fig.~\ref{fig:RTD_comparison_with_psi0_fits}(a) show 
$\Psi_0^{\scriptscriptstyle\rm fit}(t)$ 
for these parameters. 
For $\beta k L^2 = 10$, the fit closely matches the red line,
meaning that the tip-molecule interaction could go unnoticed. In
that case one would obtain
$D_0^{\scriptscriptstyle{\rm fit}} =
L^2/\pi^2\tau_L^{\scriptscriptstyle{\rm fit}} \approx 1.7\times 10^4
L^2/\rm s$, which is 70\% larger compared to the true value $D_0=10^4
L^2/\rm s$. This shows that estimates of molecule diffusion coefficients can
strongly deviate from the true value due to tip-molecule interactions,
even if an acceptable fit to experimental data under neglect of these interactions is possible.

For the larger stiffness, $\beta k L^2 = 50$, the curve $\Psi_0^{\rm fit}$ in Fig.~\ref{fig:RTD_comparison_with_psi0_fits}(a)
does not give an acceptable match with the red curve.
The RTD analysis would thus point to the presence of
tip-molecule interactions. That the dashed line
underestimates $\Psi(t)$ for short times and overestimates $\Psi(t)$
for intermediate times is an indicator that the tip-molecule
interaction is repulsive.

For the attractive interaction, we find
$\tilde{\epsilon}^{\scriptscriptstyle{\rm fit}} =
\tilde{\epsilon}_\text{max}$ even when further increasing
the upper limit $\tilde{\epsilon}_\text{max}$ of the fitting range for $\tilde{\epsilon}^{\scriptscriptstyle{\rm fit}}$. 
This is analogous to the situation
described above for the repulsive interaction, where now we have
an overestimation of $\Psi(t)$ at small times. We chose
$\tilde{\epsilon}^{\scriptscriptstyle{\rm fit}} = 0.1$.
For the scaled
stiffness $\beta k L^2=-10$, the fit (dashed line) of
Eq.~\eqref{eq:RTD0} to $\Psi(t)$ (blue line) in
Fig.~\ref{fig:RTD_comparison_with_psi0_fits}(b) gives
$\tau_L^{\scriptscriptstyle{\rm fit}}=2.0\times10^{-5}\,\rm s$. In
contrast to the repulsive case for the small stiffness, however, 
the corresponding curve $\Psi_0^{\rm fit}$
does not yield a good match with $\Psi(t)$. 
That the dashed line underestimates $\Psi(t)$ for intermediate
times is an indicator for an attractive tip-molecule interaction. For the larger scaled stiffness
$\beta k L^2 =-50$, a fit of Eq.~\eqref{eq:RTD0} to $\Psi(t)$ was not possible.

\section{Summary \& Conclusions}
In summary, we have presented a theory for including tip-molecule
interactions in the analysis of RTDs obtained from tunneling current
time series in STM experiments. For arbitrary interaction of finite
range, a general formal solution in terms of an eigenfunction
expansion is given. Explicit results are derived for an attractive and
repulsive harmonic tip-molecule interaction. Characteristic changes of
RTDs with the interaction strength are discussed and it shown how the
presence of attractive and repulsive interactions can be identified.
We have restricted our analysis here to the one-dimensional
case. Extensions of our methodology to two dimensions are possible and
will be addressed in forthcoming studies.

A very interesting point is the application of our theory to estimate
tip-molecule interaction parameters. As mentioned in the introduction,
the analysis of tunneling current time series allows one to determine
the ITD in addition to the RTD. Fitting a measured ITD to theoretical
predictions \cite{Hahne/etal:2013} yields an estimate of $D_0$, because the
ITD is not influenced by the tip-molecule interaction. With the
knowledge of $D_0$, fits of Eq.~\eqref{eq:RTD} to the RTD obtained
from the same experiment can then be carried out with respect to the
stiffness parameter $k$. This way an estimate for the sign and
strength of tip-molecule interactions can be obtained.

This opens further interesting possibilities also for measuring
molecule-molecule interactions when molecules can be picked up by an
STM tip. DFT calculations could test the reliability of such methods
in future studies.


%

\end{document}